\documentclass[twocolumn,prl,aps,showpacs,superscriptaddress,floatfix]{revtex4-1}
\usepackage[latin1]{inputenc}
\usepackage{bm}
\usepackage[usenames]{color}
\usepackage{multirow}
\usepackage{amssymb}
\usepackage{amsbsy}
\usepackage{amsmath}
\usepackage{stmaryrd}
\usepackage{graphicx}
\usepackage{epsfig}
\usepackage{placeins}
\usepackage{ulem}
\makeatletter

\newcommand{\tn}{\textnormal}

\begin{document}

\title{Thermal conductivity of the one-dimensional Fermi-Hubbard model}

\author{C. Karrasch} 
\affiliation{Dahlem Center for Complex Quantum Systems and Fachbereich Physik, Freie Universit\"at Berlin, 14195 Berlin, Germany}

\author{D. M. Kennes}
\affiliation{Institut f\"ur Theorie der Statistischen Physik, RWTH Aachen University and JARA-Fundamentals of Future Information Technology, 52056 Aachen, Germany}

\author{F. Heidrich-Meisner}
\affiliation{Department of Physics and Arnold Sommerfeld Center for Theoretical Physics,
Ludwig-Maximilians-Universit\"at M\"unchen, 80333 M\"unchen, Germany}

\date{\today}

\begin{abstract}
We study the thermal conductivity of the one-dimensional Fermi-Hubbard model at finite temperature using a 
density matrix renormalization group approach. 
The integrability of this model gives rise to ballistic thermal transport. We calculate the temperature dependence 
of the thermal Drude weight at half filling for various interactions strength.   The finite-frequency contributions originating from the fact that the energy current is not a conserved quantity are investigated as well.
We report evidence that breaking the integrability through a nearest-neighbor interaction leads to vanishing Drude weights
and diffusive energy transport.
Moreover, we demonstrate that energy spreads ballistically in local quenches with initially inhomogeneous energy density  
profiles in the integrable case. We discuss the  relevance of our results for thermalization in ultra-cold quantum gas experiments
and for transport measurements with quasi-one dimensional materials.
\end{abstract}

\maketitle

{\it Introduction.}
Improving our understanding of transport in one-dimensional (1D) strongly correlated systems (SCS) 
is  an active field
in condensed matter theory.  While in 1D, the existence of powerful numerical \cite{white92,schollwoeck05,schollwoeck11} and analytical \cite{essler-book,giamarchi} methods makes it possible to  obtain
{\it quantitative} results (see, e.g., \cite{Kluemper2000,Caux2005,Pereira2006,essler-book}), transport coefficients are very hard to come by exactly and are  challenging 
quantities to determine numerically.
Early  studies suggested that
integrable  systems such as the 1D spin-1/2 Heisenberg  or  Fermi-Hubbard model (FHM) may possess
ballistic transport properties at finite temperatures \cite{zotos97}.
In linear response theory, ballistic dynamics  manifests itself through non-zero Drude weights.
The so far best understood model is the spin-1/2 XXZ chain, for which the Drude weight for thermal transport has been calculated exactly
\cite{kluemper02,sakai03}, while substantial progress has recently been made regarding  the spin conductivity 
\cite{prosen11,herbrych11,karrasch13,prosen13,steinigeweg14,prelovsek04,steinigeweg09,sirker11,steinigeweg12,karrasch14}.

The theory of transport in the 1D FHM is much less advanced and has focussed on   spin and charge transport   \cite{kirchner99,peres00,carmelo12,prosen12,prosen14,karrasch14a,Jin2015}.
The thermal conductivity can,  by using the Kubo formula \cite{mahan,luttinger64}, be written as  
\begin{equation}\label{eq:sigma}
\mbox{Re}\, \kappa(\omega) = 2 \pi D_{\rm th}(T)\delta(\omega) + \kappa_{\rm reg}(\omega)
\end{equation}
with the  thermal Drude weight $D_{\rm th}(T)$ and a regular finite-frequency contribution $ \kappa_{\rm reg}(\omega)$.
Since in SCS, the Wiedemann-Franz law is not necessarily  valid, independent calculations of 
charge  and thermal transport are required.

The formal argument to prove a nonzero Drude weight  relies on the  Mazur inequality \cite{suzuki71,mazur69,zotos97} 
\begin{equation}\label{eq:bound}
D_{\rm th} \geq \frac{1}{2T^2 L}\sum_i \frac{\langle Q_i I_{\rm th}\rangle^2}{\langle Q_i^2\rangle}\,,
\end{equation}
where $I_{\rm th}$ is the energy-current operator and the $Q_i$ are local or quasi-local conserved quantities  \cite{prosen11,prosen13,mierzejewski14}. 
In the presence of interactions, nontrivial $Q_i$ leading to finite Drude weights typically  exist in integrable models \cite{zotos97}.
 For instance, for the spin-1/2 XXZ chain, 
the 
energy current $I_{\rm th}=Q_3$ itself is conserved (implying that $\kappa_{\rm reg}(\omega) =0$), 
while for the FHM, $I_{\rm th}$ only has a partial overlap with $Q_3$ (both operators have a similar structure \cite{suppmat}) such that 
while $D_{\rm th}>0$, also
$\kappa_{\rm reg}(\omega)\not=0$ \cite{zotos97}. 
As a consequence, the half-filled FHM realizes an unusual behavior:  ballistic thermal transport \cite{zotos97},
yet  diffusive charge conduction \cite{karrasch14a,Jin2015} at temperatures $T>0$.

Here,  we  address the outstanding problem of {\it quantitatively} calculating Re\,$\kappa(\omega)$ at $T>0$ for the FHM at half filling   by using a finite-temperature density matrix renormalization group (DMRG) method \cite{karrasch12,karrasch13a,barthel13,verstraete04,feiguin05,schollwoeck11},  previously applied to  both charge transport in the FHM \cite{karrasch14a} and 
transport in quasi-1D spin-1/2 systems \cite{karrasch12,karrasch13,karrasch14,deluca14,karrasch15}. 
We obtain the energy-current autocorrelation functions 
$C_{\rm th}(t) = \tn{Re}\,\langle I_{\rm th} (t) I_{\rm th}\rangle/L\,$ 
 from time-dependent simulations. 
Since $C_{\rm th}(t)$ saturates fast at a time-independent non-zero value, we are able to extract (i) the thermal Drude weight and (ii) the regular part from 
a Fourier transformation in combination with linear prediction \cite{barthel09}. Moreover, we consider the extended FHM as an example for a non-integrable model
and provide evidence that 
 ballistic contributions are absent, with a diffusive form of the low-frequency $\kappa_{\rm reg}(\omega)$.

  The FHM has been realized with ultra-cold quantum gases \cite{schneider08,joerdens08,hart2015,edge2015,Omran2015,haller2015,greif2016,boll2016,Cocchi2016,Cheuk2016}. In ultra-cold quantum gases, 
relaxation processes play an important role
for reaching thermal equilibrium during the state preparation \cite{hung10,schmitt13} and thermometry is an open experimental problem \cite{mckay}.  Furthermore, understanding thermalization dynamics and non-equilibrium transport as such have been  the goal of several optical-lattice experiments with Hubbard systems \cite{schneider12,trotzky12,ronzheimer13,pertot14}. We demonstrate that real-space perturbations in the energy density spread ballistically in the
1D FHM at $T>0$ while charge diffuses \cite{karrasch14a,Jin2015}, providing a route to experimentally observing the 
qualitative difference between charge and energy dynamics in this model. 

{\it Definitions.}
The Hamiltonian of the extended FHM is given by  $H=\sum_{l=1}^{L-1}h_l$ with local terms 
\begin{eqnarray}\label{eq:h}
h_l&= &-{t_0}  \sum_\sigma \left(c_{l\sigma}^\dagger c_{l+1\sigma}^{\phantom{\dagger}} + \tn{h.c.} \right) +V (n_{l}-1)(n_{l+1}-1) \\ 
&+& \frac{U}{2}\left((n_{l\uparrow}-\frac{1}{2})(n_{l\downarrow}-\frac{1}{2})+(n_{l+1\uparrow}-\frac{1}{2})(n_{l+1\downarrow}-\frac{1}{2})\right) 
\nonumber\,,
\end{eqnarray}
where $c_{l\sigma}$ annihilates a fermion with spin $\sigma$ on site $l$, and $n_{l\sigma}=c_{l\sigma}^\dagger c_{l\sigma}^{\phantom{\dagger}}$.   $U$ and $V$ denote the onsite and the nearest-neighbor Coulomb repulsion, respectively. We use open boundary conditions. All results in the main text are for half filling $n=N/L=1$, where $N$ is the total number of fermions.
For convenience, we implemented the FHM as a two-leg spin-1/2 ladder \cite{suppmat}.

We derive the energy current from the continuity equation \cite{zotos97}, leading to $I_\tn{th} = i \sum_{l=1}^{L-2} [h_l,h_{l+1}]$ 
(for the full expression, see \cite{suppmat}). 
At $n=1$, particle-hole symmetry leads to a vanishing thermopower \cite{beni75,suppmat} and thus,
the thermal conductivity stems solely from energy-current correlations.

\begin{figure}[t]
\includegraphics[width=0.9\columnwidth]{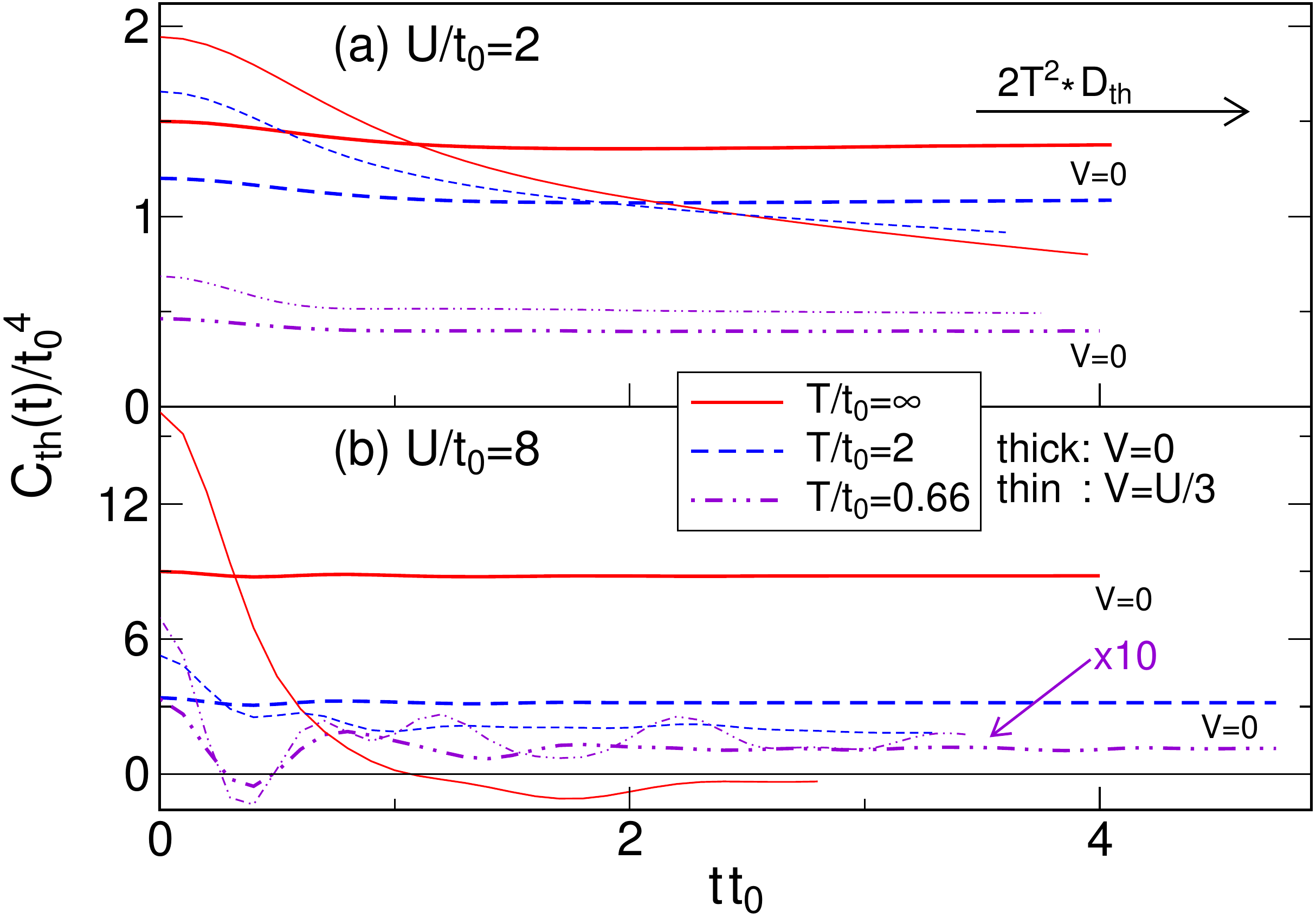}
\caption{(Color online) DMRG results for $C_{\rm th}(t)$   for various $T>0$ at (a) $U/t_0=2$ and (b) $U/t_0=8$. 
 Thick lines:  $V=0$, thin lines: $V= U/3$. The curves for $U/t_0=8$, $T/t_0=0.66$ are multiplied  by a factor of 10.
}\label{Fig1}
\end{figure}

{\it Numerical method.}
The thermal Drude weight is related to the long-time asymptote of the current correlation functions via
\begin{equation}\label{eq:dw}
D_\tn{th} = \lim_{t\to\infty}\lim_{L\to\infty} \frac{C_\tn{th}(t)}{2T^2}\,,
\end{equation}
and the regular part of the conductivity defined in Eq.~(\ref{eq:sigma}) can be obtained from ($\tilde C_\tn{th}(t) = C_\tn{th}(t) - 2T^{2}D_\tn{th}$)
\begin{equation}\label{eq:sigma1}\begin{split}
&\kappa_\tn{reg}(\omega) =\frac{1-e^{-\frac{\omega}{T}}}{\omega T }
\tn{Re} \hspace*{-0.1cm}\int_0^{\infty}\hspace*{-0.2cm}dte^{i\omega t} \hspace*{-0.1cm}\lim_{L\to\infty}\tilde C_\tn{th}(t) .
\end{split}\end{equation}
Note that the derivation of the Kubo formula for $\kappa$ is more subtle than for the charge conductivity (see, e.g., Refs.~\cite{luttinger64,Kubo1957b,Kubo1957a,Luttinger1964,louis03,gemmer06} and references therein and \cite{suppmat} for a discussion),
while there is also ongoing research on thermal and energy transport in open quantum system  (see, e.g., \cite{michel05,prosen09,arrachea09,mendoza-arenas13}).

Our finite-T DMRG method, implemented via matrix-product states \cite{fannes91,ostlund91,verstraete06,verstraete08}, 
is based on the purification trick \cite{verstraete04a} (see \cite{white09,barthel09,zwolak04,sirker05,white09,barthel13} for related work).
Thus, we simulate 
pure states that live in a Hilbert space spanned by the physical  and auxiliary (ancilla) degrees of freedom. Mixed states are obtained by tracing over
the ancillas. 
In order to access time scales as large as possible, we employ a finite-temperature disentangler \cite{karrasch12}, using  
that  purification is not unique to slow down the entanglement growth. Moreover, we exploit `time-translation invariance' \cite{barthel13},  rewrite $\langle I_\tn{th}(t)I_\tn{th}(0)\rangle=\langle I_\tn{th}(t/2)I_\tn{th}(-t/2)\rangle$, and carry out two independent calculations for $I_\tn{th}(t/2)$ as well as for $I_\tn{th}(-t/2)$. 
Since the energy current is a six-point function, a tMDRG simulation for the energy-current autocorrelation is much more demanding than it is in the charge case.
Our calculations are performed with $L=100$ sites (see Fig.~S1 in \cite{suppmat} for an analysis of the  $L$-dependence). 
The `finite-time' error of $\kappa_\tn{reg}(\omega)$  can be assessed following \cite{karrasch15},   resulting in the  error bars  shown in the figures.

\begin{figure}[t]
\includegraphics[width=0.9\columnwidth]{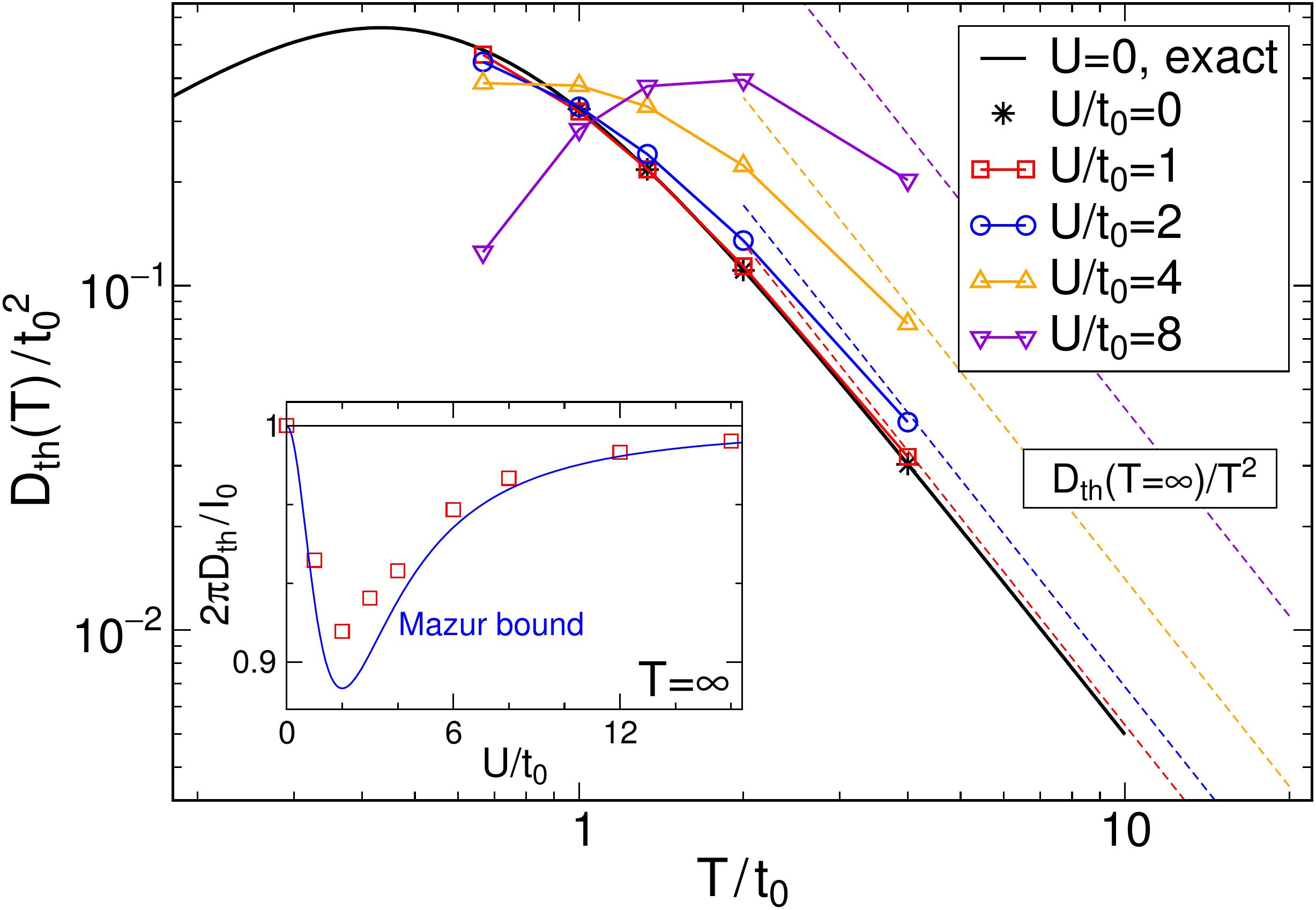}
\caption{(Color online) Thermal Drude weight of the integrable model versus  temperature for  $U/t_0=0,1,2,4,8$. 
Solid line: exact result Eq.~(\ref{eq:u0}) for  $U=0$, in excellent agreement with DMRG data. Dashed lines: High-$T$ behavior
$D_{\rm th}= D_{\rm th}^{\infty}/T^2$, with $D_{\rm th}^{\infty}$ computed using DMRG. Inset:  $2\pi D_{\rm th}/I_0$  at $T=\infty$ (squares), compared to
the lower Mazur bound \cite{zotos97} (solid line).
}\label{Fig2}
\end{figure}

{\it Real-time decay of $C_{\rm th}$.} Typical DMRG results for $C_{\rm th}(t)$ are shown in Figs.~\ref{Fig1}(a) and (b) for $U=2t_0$ and $U=8t_0$, respectively, and temperatures $T=\infty,2t_0,2t_0/3$.
We are able to reach times $t t_0\lesssim 5$. For $V=0$ (thick lines), i.e., in the integrable case, $C_{\rm th}(t)$ rapidly saturates at a constant non-zero value, reflecting the ballistic nature
of energy transport in this model.  The transients are surprisingly short compared to  spin transport in the spin-1/2 XXZ chain \cite{karrasch13} and exhibit oscillations with 
a fairly small amplitude. To illustrate the behavior in the non-integrable extended FHM, we present data for $V=U/3$ (thin lines), for which our system  Eq.~\eqref{eq:h} is still
 in the Mott-insulating phase \cite{jeckelmann02}. The DMRG results unveil a much stronger decay of $C_{\rm th}(t)$ compared to the integrable case, yet also longer transient dynamics before the asymptotic regime is reached [see, e.g., the data for $T=\infty$ shown in Fig.~\ref{Fig1}(a)]. For $U/t_0=8$ and $T=\infty$, the real-time decay of $C_{\rm th } $ is consistent with a vanishing $D_{\rm th}(T)$, as expected for this non-integrable model \cite{zotos96,hm03,zotos04,jung06,karrasch15}.

{\it Thermal Drude weight for $V=0$.} The fast saturation of $C_{\rm th}(t)$ at a constant and non-zero value  allows us to extract the
temperature dependence of $D_{\rm th}(T)$, displayed in Fig.~\ref{Fig2} for $U/t_0=0,1,2,4,8$ (note the log-log scale). 
For $U=0$, we compare our data to the exact result \cite{hm03}
\begin{equation}\label{eq:u0}
D_\tn{th}(T) = \frac{t_0^2}{2\pi T^2}\int_{-\pi}^\pi [\epsilon_k v_k f(\epsilon_k)]^2 e^{\epsilon_k/T}dk\,,
\end{equation}
where $\epsilon_k=-2t_0\cos(k)$, $v_k=-\partial\epsilon_k/\partial k$, and $f(\epsilon)=1/(1+e^{\epsilon/T})$. 
The agreement is excellent.
In general, $D_{\rm th}(T)$ has a maximum at a $U$-dependent  temperature that shifts to larger temperature as $U$ increases. In the high-temperature regime $T>t_0$,
$D_{\rm th}=D_{\rm th}^{\infty} /T^2$ (dashed lines in the figure), where the prefactor $D_{\rm th}^{\infty}$ has been extracted from the numerical data at $\beta=0$.
In the supplemental material \cite{suppmat}, we  compare the thermal Drude weight of the FHM to the one of the Heisenberg chain \cite{kluemper02}.
The latter describes the low-temperature contribution of {\it spin excitations} to the full $D_{\rm th}$ of the FHM for $U\gg t_0$ at $n=1$, while we obtained results
for the spin-incoherent regime $T\gg 4t_0^2/U$, where charge excitations dominate.

We next study how much of the full spectral weight of Re$\,\kappa(\omega)$ is in the Drude peak
by plotting  $2\pi D_{\rm th}/I_0 $ versus $U/t_0$ at $\beta =0$ in the inset of Fig.~\ref{Fig2}, where
$I_0=\int d\omega \, \mbox{Re}\, \kappa(\omega) 
$.
The Drude weight contains the full weight 
$I_0$ only at $U=0$ and for $U/t_0\to \infty$. In the former case, this results from the exact conservation of the thermal
current in the non-interacting case, while in the latter case, it is a consequence of a full suppression of any scattering 
between subspaces with different numbers of doublons as $U/t_0$ diverges.
For a finite $0<U/t_0 < \infty$, $2\pi D_{\rm th}/I_0 <1$ and it takes a minimum with $2\pi D_{\rm th}/I_0 \approx 0.92$  close to $U=2t_0$,
implying that at $n=1$, the dominant contribution to  Re$\,\kappa(\omega)$  always comes from the Drude weight.
Ref.~\cite{zotos97} provides a nonzero lower bound for $D_{\rm th}$ at $T=\infty$  by considering only $Q_3$ (a close relative of $I_{\rm th}$ \cite{suppmat}) in Eq.~\eqref{eq:bound}.
By comparison to this lower bound  (solid line in the inset of Fig.~\ref{Fig2}), we conclude that the
position of this minimum can be understood from the competition of the $U$-dependent and $U$-independent contributions to the Drude weight and to the
total weight $I_0 \propto \langle I_{\rm th} ^2\rangle $.
Moreover, the lower-bound from \cite{zotos97} is not exhaustive (see also Fig.~S2 in \cite{suppmat} showing $D_{\rm th}$ and the lower bound as a function of $n$).

\begin{figure}[t!]
\includegraphics[width=0.9\columnwidth]{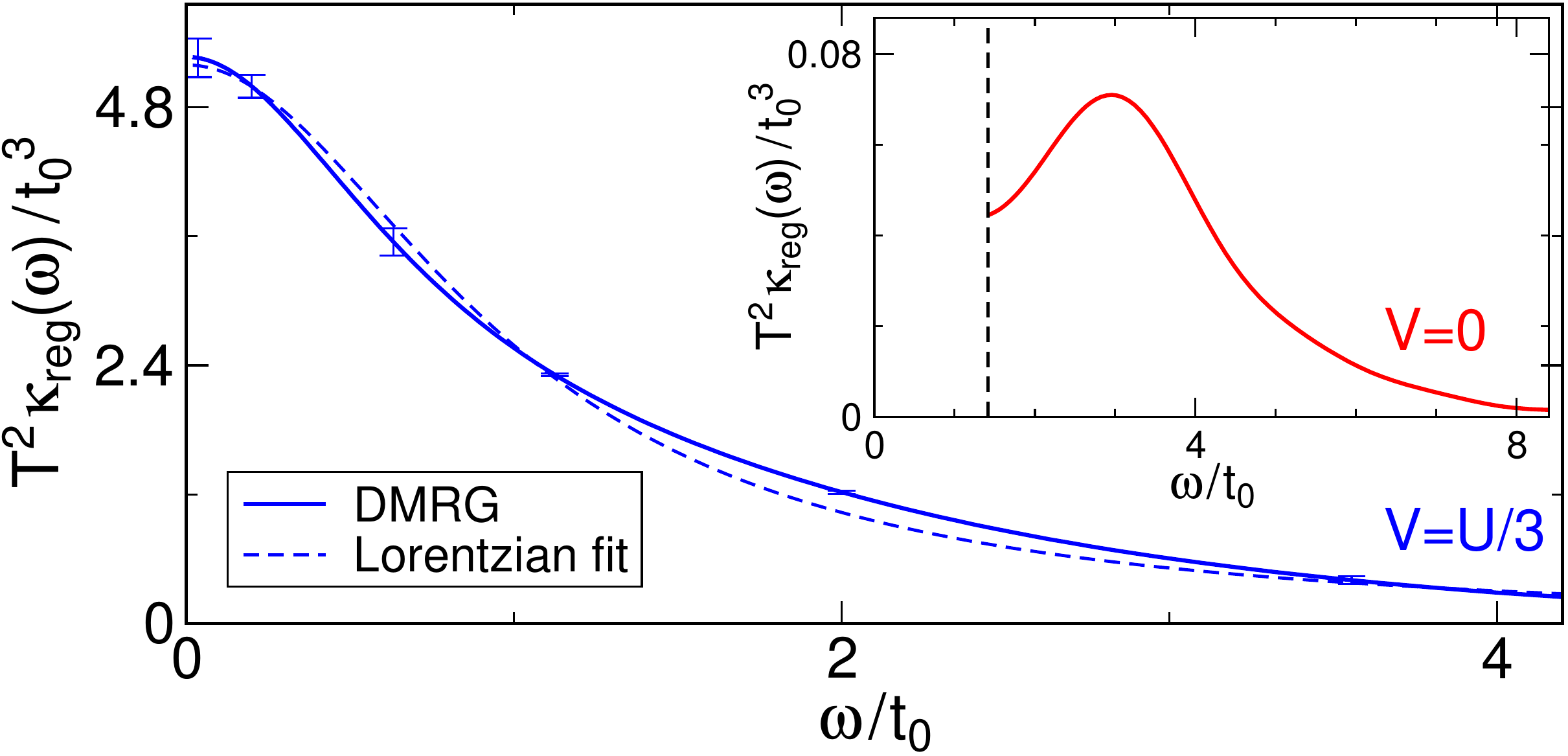}
\caption{(Color online) Regular part $\kappa_{\rm reg}(\omega)$ for $U/t_0=4$ at $T=\infty$.  Main panel: non-integrable case ($V=U/3$).  
Dashed line: Fit of DMRG data to a Lorentzian. Inset:  Integrable case ($V=0$). The data are shown only for $\omega/t_0\gtrsim1.4$,
whereas for smaller  frequencies, uncertainties become too large (see the text). 
}\label{Fig3}
\end{figure}

{\it Non-integrable model and low-frequency dependence of $\kappa_{\rm reg}(\omega)$.} Upon breaking integrability, our results for $C_{\rm th}(t)$ indicate a vanishing Drude weight, at least at 
high temperatures and for intermediate values of $U/t_0$. This raises the question of the functional form of $\kappa_{\rm reg}(\omega)$
for $V\not=0$. Figure~\ref{Fig3}  shows $\kappa_{\rm reg}(\omega)$ for $U/t_0=4$ at $T=\infty$ and $V=U/3$ (main panel) and 
$V=0$ (inset). In the non-integrable case, $\kappa_{\rm reg}(\omega)$ has a broad peak at zero frequency, which is very close to a Lorentzian
(the dashed line is a fit to the data). This demonstrates that  standard diffusion is realized in the extended Hubbard model. 
In the integrable case, we often observe maxima in $\kappa_{\rm reg}(\omega)$ at $\omega>0$, which seem to be related to the charge gap.
Due to the uncertainties involved in extracting the frequency dependence, which are due to the finite times reached in the simulations and the extraction of the 
Drude weight, we are not able to resolve the low-frequency regime for $V=0$. Therefore, the question of whether $\kappa_{\rm reg}(\omega\to 0)$
is zero or finite in the integrable case, which has been intensely studied for spin transport in the spin-1/2 XXZ chain \cite{sirker09,grossjohann10,sirker11,herbrych12,karrasch15,brenig15},
remains an open problem.

\begin{figure}[t]
\includegraphics[width=0.48\columnwidth]{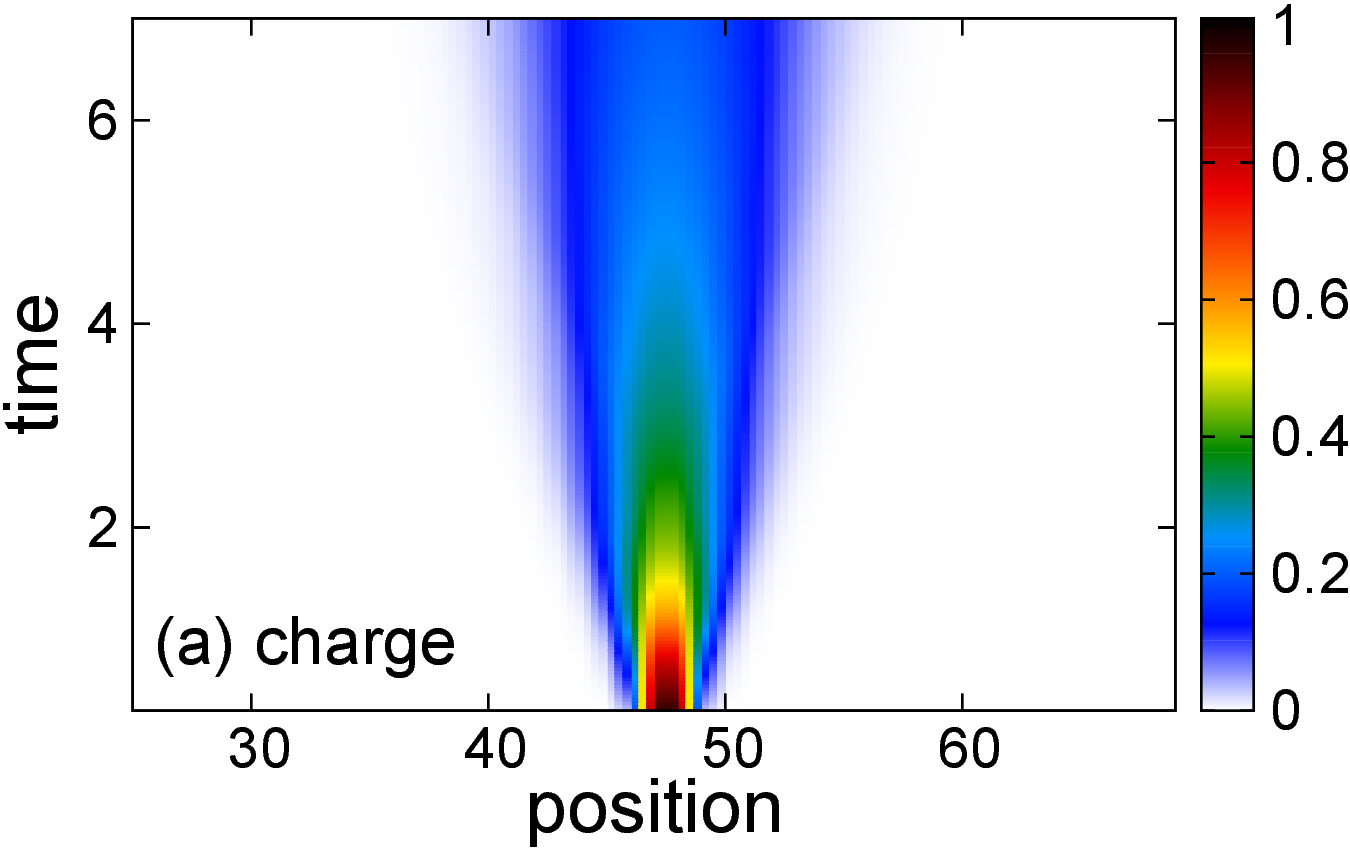}\hspace*{0.03\columnwidth}
\includegraphics[width=0.48\columnwidth]{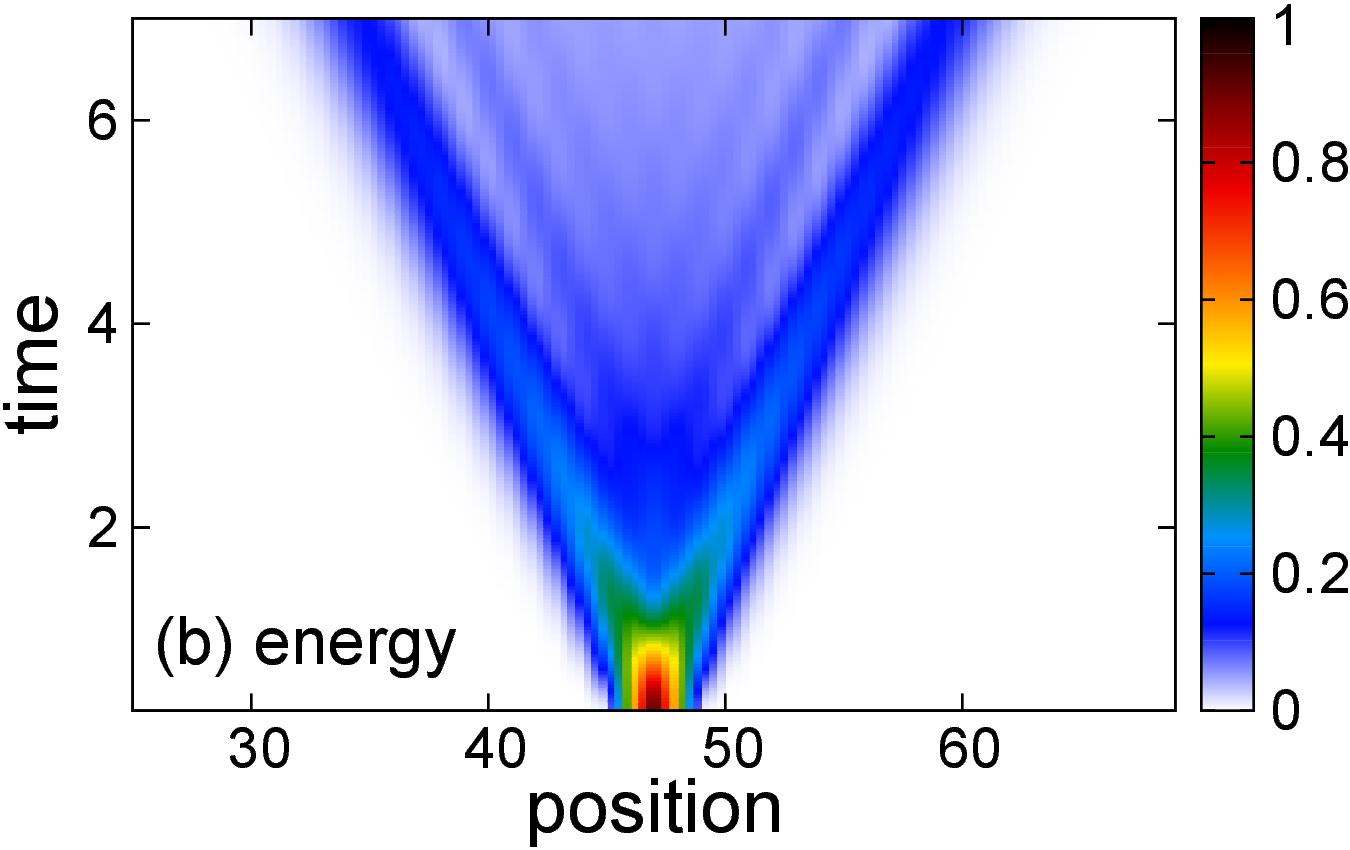} \\[1ex]

\includegraphics[width=0.49\columnwidth]{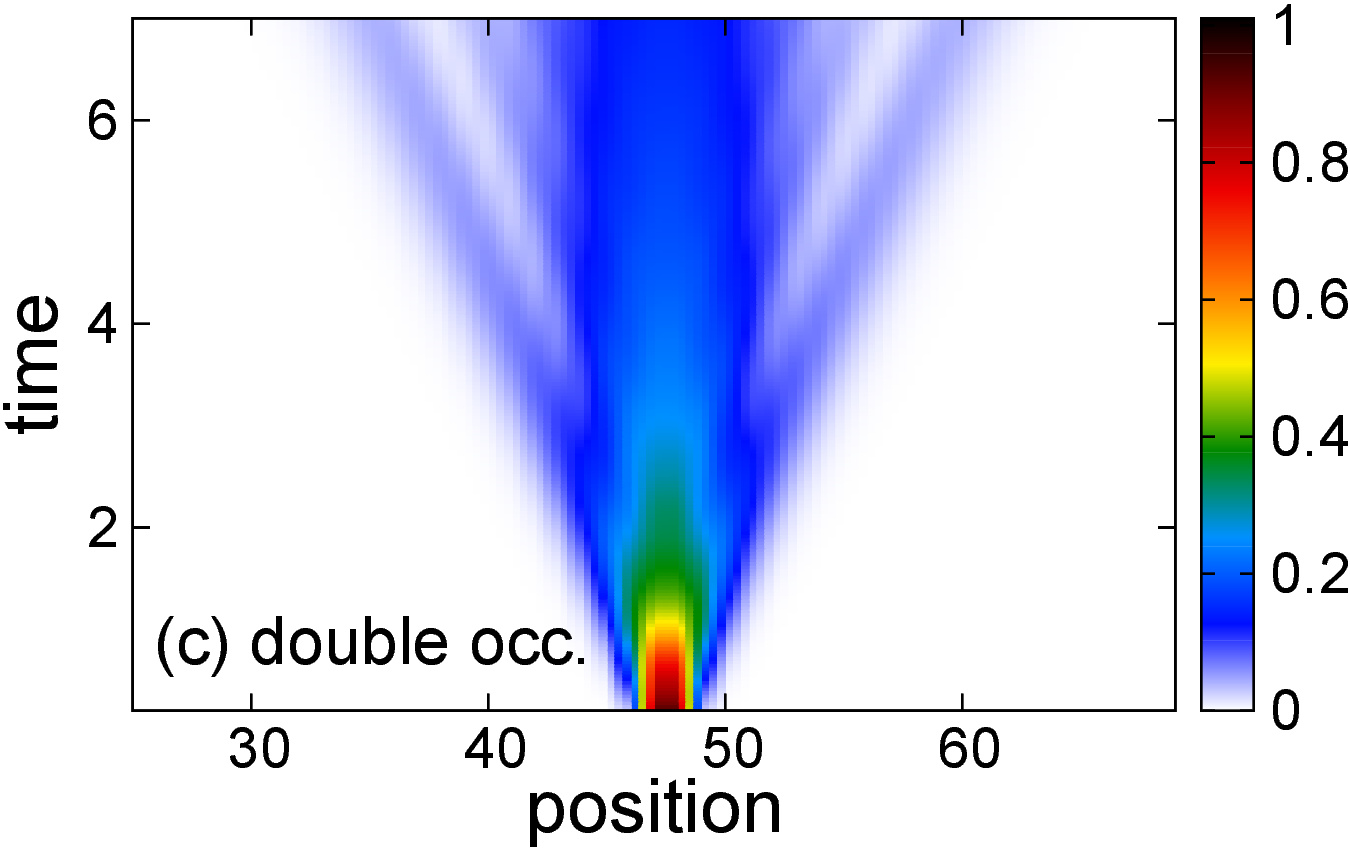}\hspace*{0.02\columnwidth}
\includegraphics[width=0.48\columnwidth]{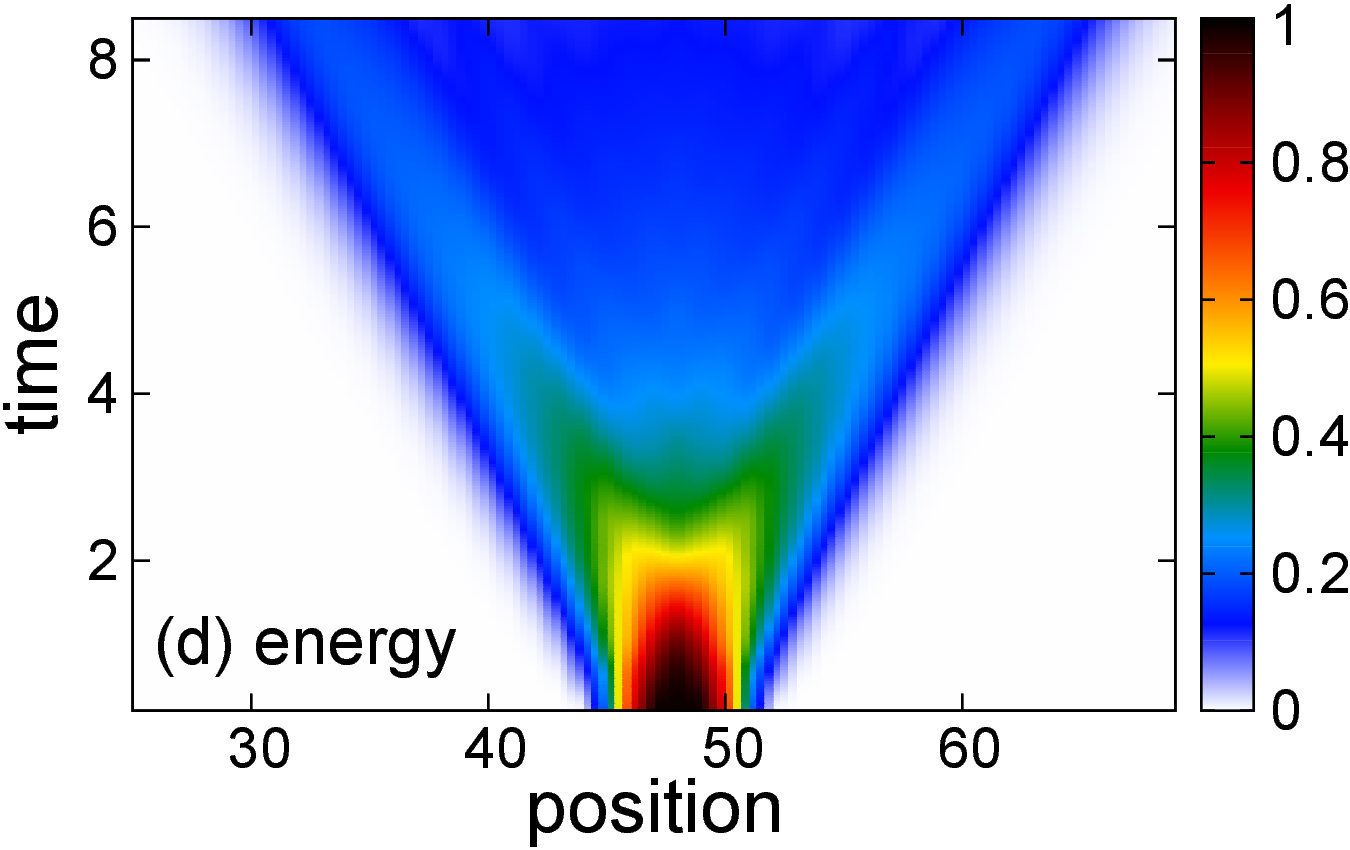}
\caption{(Color online) Local quenches for  $U/t_0=8$ and $V=0$. (a--c) 
At time $t=0$, a local excitation $(|0\rangle+|\hspace*{-0.1cm}\uparrow\rangle+|\hspace*{-0.1cm}\downarrow\rangle+1.1|\hspace*{-0.1cm}\uparrow\downarrow\rangle)$ is prepared at two sites in the center of an otherwise equilibrated system at temperature $T=\infty$.
 Real-time evolution of (a) the normalized local charge density $[\langle n_{l}(t)\rangle-1]/n_0$, (b) the normalized local energy density $\langle h_l(t)\rangle/E_0$, and (c) the relative probability for double occupancy $[P(n_l=2)-1/4]/p_0$. The normalization constants read $n_0=(2*1.1^2+2)/(1.1^2+3)-1, E_0\approx0.025t_0$, and $p_0=1.1^2/(1.1^2+3)-1/4$.
(d) $T$-quench ($T_1/t_0=10$, $T_2=\infty$):  relative energy density $\langle h_l(t)\rangle/(-0.245t_0)$.
}\label{Fig4}
\end{figure}

{\it Spreading of local perturbations.} The presence of a ballistic contribution in the linear response functions
translates into the ballistic spreading of perturbations in the local density \cite{polini07,steinigeweg09,langer09,langer11,jesenko11,karrasch14,yonghong}.
To illustrate this connection, we study a Hubbard chain at infinite temperature and introduce a perturbation in the local charge density at
$t=0^+$. This also causes a perturbation in the energy density. We  measure the time evolution of both densities $\rho_{{\rm ch},l}(t)=\langle n_l(t) \rangle$
and $\rho_{\rm th, l}(t)=\langle h_l(t) \rangle$,  presented in Figs.~\ref{Fig4}(a) and (b). While the energy density 
shows the typical features of a 
ballistic dynamics \cite{langer11,karrasch14}, the charge density exhibits a much slower spreading and does not form fast ballistic
jets, nicely illustrating  the different nature of energy versus charge transport in this model.
To become more quantitative, we compute the  spatial variances associated with the density $\rho_{{\rm th, ch},l}(t)$
\begin{equation}
\sigma^2_{\rm th,ch}(t) = \frac{1}{\mathcal{N}_{\rm th,ch}}\sum_{l=n_0}^{L-n_0} (l-l_0)^2 \big[\rho_{{\rm th, ch},l}(t)-\rho_{{\rm th, ch}}^\tn{bg}\big]\,
\end{equation}
where $l_0$ is the center of the wave packet, $n_0$ cuts off boundary effects, $\rho_{{\rm th, ch}}^\tn{bg}$ denotes the bulk background density, and $\mathcal{N}_{\rm th,ch}$ is the excess particle number or energy induced by the wave packet. As expected we find $\delta\sigma_{\rm th}^2 = \sigma_{\rm th}^2(t)-\sigma_{\rm th}^2(t=0)\propto t^2$ yet a much slower
growth for the charge $\sigma^2_{\rm ch}\propto t^\alpha$ with $1/2<\alpha < 1$ [see Fig.~S4]. The determination of the exact exponent would require longer times
and is related to the low-frequency behavior of the charge conductivity, yet clearly, charge dynamics is not ballistic. Another illustration for the ballistic energy spreading can be obtained in $T$-quenches \cite{karrasch14}, in which we embed a region with $T_2$ into a larger system that is at $T_1<T_2$, which overall has a homogeneous spin and charge density. 
An example is shown in Fig.~\ref{Fig4}(d) and as expected, the variance of this wave-packet grows as $\delta\sigma_\tn{th}(t)^2\propto t^2$, illustrating that energy spreads ballistically
in the temperature quench as well.

The time evolution of the double occupancy (accessible in optical-lattice experiments) $ d(t)=\langle n_{i\uparrow} n_{i\downarrow}\rangle(t)$ 
is shown in Fig.~\ref{Fig4}(c) for the quench of Figs.~\ref{Fig4}(a) and (b). 
The profile exhibits fast ballistic jets and the associated variance $\sigma_d^2(t) =\frac{1}{D} \sum_l (l-l_0)^2[\langle d_l(t)\rangle-d^\tn{bg}]$ ($D = \sum_l \langle d_l \rangle $)
increases approximately quadratically  at long times and is thus sensitive to the fast spreading of the energy density. 
For $V\not= 0$ (see Fig.~S5 for an illustration), the variances of both energy and double occupancy increase much slower than quadratically in time.
As a consequence of the ballistic energy transport in a 1D FHM we expect the
absence of thermalization in related quantum gas experiments.
From the long-time behavior of the respective width $\delta\sigma_\tn{th,ch}(t)$, one can extract diffusion constants \cite{steinigeweg09,karrasch14} or Drude weights (see, e.g., \cite{yan2015})
via Einstein relations (as we verified for our case), providing an experimental means of measuring transport coefficients.
Such a local, real-space and real-time  probe for thermal transport has recently been used in experiments with  low-dimensional quantum magnets
\cite{otter09}. Given that a coupling to phonons cannot be avoided in quantum magnets \cite{hess07,montagnese2013},
quantum-gas microscope experiments \cite{edge2015,Omran2015,haller2015,greif2016,boll2016,Cocchi2016,Cheuk2016} could provide a means
of studying energy and charge transport in the FHM, which is easier to realize with ultracold quantum gases than the spin-1/2 XXZ chain in its massive
regime, where a similar coexistence of diffusive spin transport \cite{karrasch14,steinigeweg12} and ballistic energy transport exists \cite{zotos97,hm02,sakai03}.

{\it Summary and outlook.}
We computed the thermal conductivity of the 1D FHM using a finite-$T$ DMRG method.  We confirm the ballistic nature of 
thermal transport in the integrable case and we studied
the temperature dependence of the Drude weight.
The lower bound for $D_{\rm th}$ from \cite{zotos97} is not exhaustive implying that more local (or even quasi-local \cite{prosen11,prosen13,mierzejewski14}) conserved quantities
than just $Q_3$ play a role. 
We further demonstrated that 
the coexistence of diffusive charge transport and ballistic thermal transport is directly reflected in local quantum quench dynamics, presumably accessible
  to fermionic quantum gas microscopes \cite{edge2015,Omran2015,haller2015,greif2016,boll2016,Cocchi2016,Cheuk2016}. 
For the extended Hubbard model, we identified  regimes in which first,  
the Drude weight  clearly vanishes as system size increases and second,  the low-frequency dependence
is compatible with diffusive dynamics. 

From the theoretical point of view, 
an exact calculation of $\kappa(\omega)$
exploiting the integrability of the model constitutes an open problem.
In view of the existence of quasi-1D materials described by the (extended) Hubbard model
 (including some Beechgard salts \cite{jerome,vescoli},  organic materials  \cite{hasegawa97,claessen04,wall11}, anorganic systems
such as Sr$_{2}$CuO$_{3}$ \cite{ono} and carbon nanotubes \cite{bockrath99,ishii03,deshpande09}), a detailed analysis of energy transport and the 
calculation of diffusion constants  is  desirable.
Finally, the investigation  of thermoelectric effects in SCS are  a timely topic (see, e.g., \cite{peterson07,shastry09,wissgott10,hong13,Mierzejewski2014,kokalj14})
 and should be feasible with our technique, at least at high temperatures.

{\it Acknowledgment.} We thank C. Hess, {T. Prosen}, {R. Steinigeweg}, and X. Zotos for very helpful discussions. We thank M. Medenjak and T. Prosen for pointing out a mistake in a previous version of Fig. S2 to us. 
 We acknowledge support by the Nanostructured Thermoelectrics program of LBNL (C.K.) as well as by the DFG through the
Research Training Group 1995 (D.M.K.) and the Emmy Noether program (C.K.).

\newpage

\setcounter{figure}{0}
\setcounter{equation}{0}

\renewcommand{\thetable}{S\arabic{table}}
\renewcommand{\thefigure}{S\arabic{figure}}
\renewcommand{\theequation}{S\arabic{equation}}

\renewcommand{\thesection}{S\arabic{section}}

\subsection*{S1.~Energy current operator in fermionic and spin language}

\begin{figure}[t]
\includegraphics[width=0.9\columnwidth]{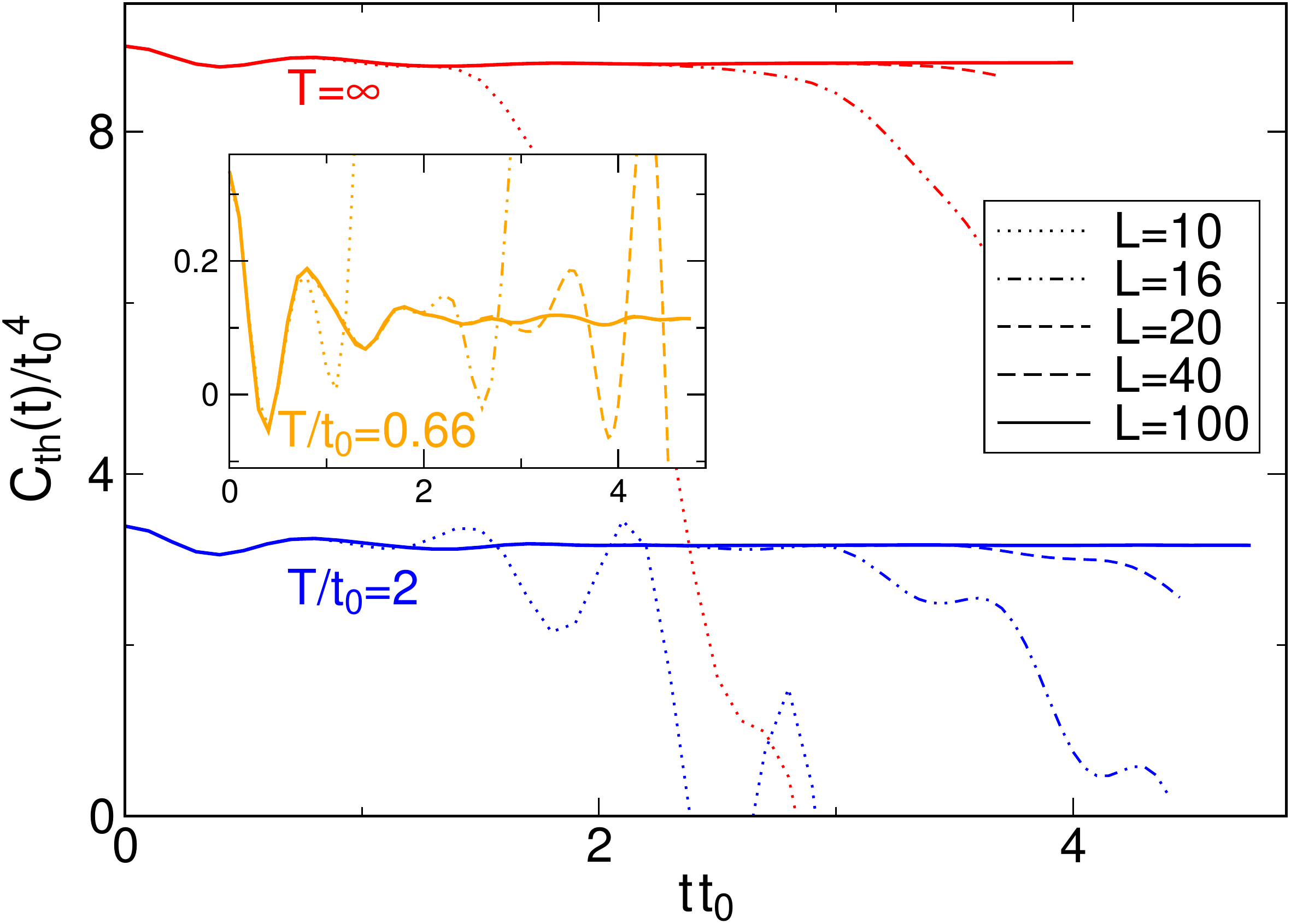}
\caption{(Color online) Energy-current autocorrelation function at $U/t_0=8$ and $V=0$ calculated for various system sizes $L$. Note that the curves for $L=40$ and $L=100$ are indistinguishable.  
}\label{FigSupL}
\end{figure}

Following Ref.~\onlinecite{zotos97}, we define the energy current operator using a continuity equation. This yields $I_\tn{th} = i \sum_{l=1}^{L-2} [h_l,h_{l+1}]$. For $V=0$, this results in 
\begin{equation}\begin{split}
 I_\tn{th} = \sum_{l,\sigma}t_0^2\Big[&\big(i c_{l+1\sigma}^\dagger c_{l-1\sigma}^{\phantom{\dagger}}+\tn{h.c.}\big)\\
& -\frac{U}{2}\big(j_{l-1\sigma}+j_{l\sigma}\big)\big(n_{l\bar\sigma}-\frac{1}{2}\big)\Big]\,,
\end{split}\end{equation}
where $j_{l\sigma}=it_0c_{l+1\sigma}^\dagger c_{l\sigma}^{\phantom{\dagger}}+\tn{h.c.}$ is the local charge current. 
$I_{\rm th}$ has the same structure as the conserved charge $Q_3$, except for a different prefactor of $1/2$ in front of the $U$-dependent term \cite{zotos97}.

Within the DMRG numerics, we implement a spin version of the Hamiltonian as well as of $I_\tn{th}$, which we obtain via a Jordan-Wigner transformation:
\begin{equation}
c_{l\uparrow} = (\sigma_1^z\cdots\sigma_{l-1}^z)\sigma_l^-,~
c_{l\downarrow} = \Big(\prod_{l=1}^L\sigma_l^z\Big)(\tau_1^z\cdots\tau_{l-1}^z)\tau_l^-~,
\end{equation}
where $\sigma_l^{x,y,z}$ and $\tau_l^{x,y,z}$ denote standard Pauli matrices, and $\sigma_l^\pm = (\sigma_l^x\pm i\sigma_l^y)/2$, $\tau_l^\pm = (\tau_l^x\pm i\tau_l^y)/2$. This leads to (for notational simplicity we again set $V=0$)
\begin{equation}
 H = \sum_{l} \Big[t_0\big(\sigma^+_{l+1}\sigma^-_l + \tau^+_{l+1}\tau^-_l + \tn{h.c.}\big)+\frac{U}{4}\sigma^z_l\tau^z_l\Big]\,,
\end{equation}
and
\begin{equation}\begin{split}
 I_\tn{th} = \sum_{l}&\Big\{\big(-it_0^2\sigma^+_{l+1}\sigma^z_l\sigma^-_{l-1}-it_0^2\tau^+_{l+1}\tau^z_l\tau^-_{l-1} +\tn{h.c.}\big)\\
 & -\frac{U}{4}\Big[\big(j^\sigma_{l-1}+j^\sigma_{l}\big)\tau^z_l +\big(j^\tau_{l-1}+j^\tau_{l}\big)\sigma^z_l \Big]\Big\}\,,
\end{split}\end{equation}
where $j_l^\sigma=-it_0\sigma^+_{l+1}\sigma^-_l+\tn{h.c.}$, $j_l^\tau=-it_0\tau^+_{l+1}\tau^-_l+\tn{h.c.}$.

\begin{figure}[t]
\includegraphics[width=0.9\columnwidth]{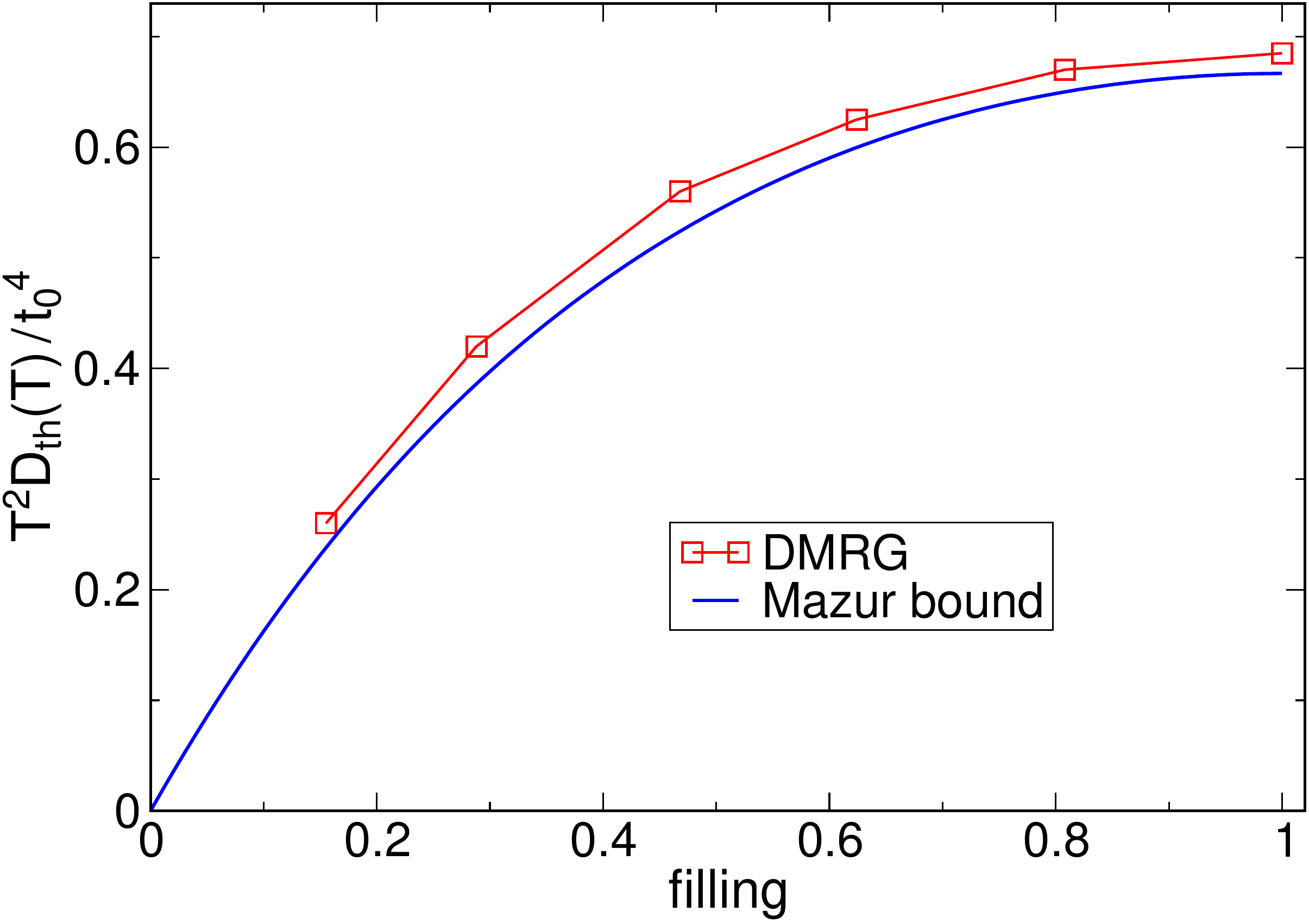}
\caption{(Color online) Filling-dependence of the Drude weight at $U/t_0=2$ and $T=\infty$. The lower Mazur bound from Ref.~\onlinecite{zotos97} is shown as a comparison.
}\label{FigSupn}
\end{figure}

\subsection*{S2.~Transport coefficients}

In general, the heat current $I_q$ of an electronic system is defined as \cite{mahan,ashcroft} 
\begin{equation}
I_q = I_{\rm th} - \mu I\,.
\end{equation}
As a consequence, one needs to consider a two-by-two matrix of transport coefficients $L_{ij} $ that relate
external forces to currents. One possible choice is to work with the heat current $I_q$ and the particle(electrical) current $I$ and the
corresponding forces $F_J=-\frac{1}{T}\nabla (\mu +V)$, where $V$ is the electrostatic potential, $\mu$ is the chemical potential  and $F_q=\nabla (1/T)$, resulting in 
\begin{equation}
\left( \begin{array}{c}  I \\ I_q \end{array}\right) =  \left( \begin{array}{cc} L_{11} & L_{12} \\ L_{21} & L_{22}\end{array}\right)  
\left( \begin{array}{c}  F_J  \\ F_{q} \end{array}\right)\,
\end{equation}
(we set the electrical charge to one for simplicity).
The transport coefficients $L_{ij}$ are computed from the respective Kubo formulae \cite{mahan}. 

\begin{figure}[t]
\includegraphics[width=0.9\columnwidth]{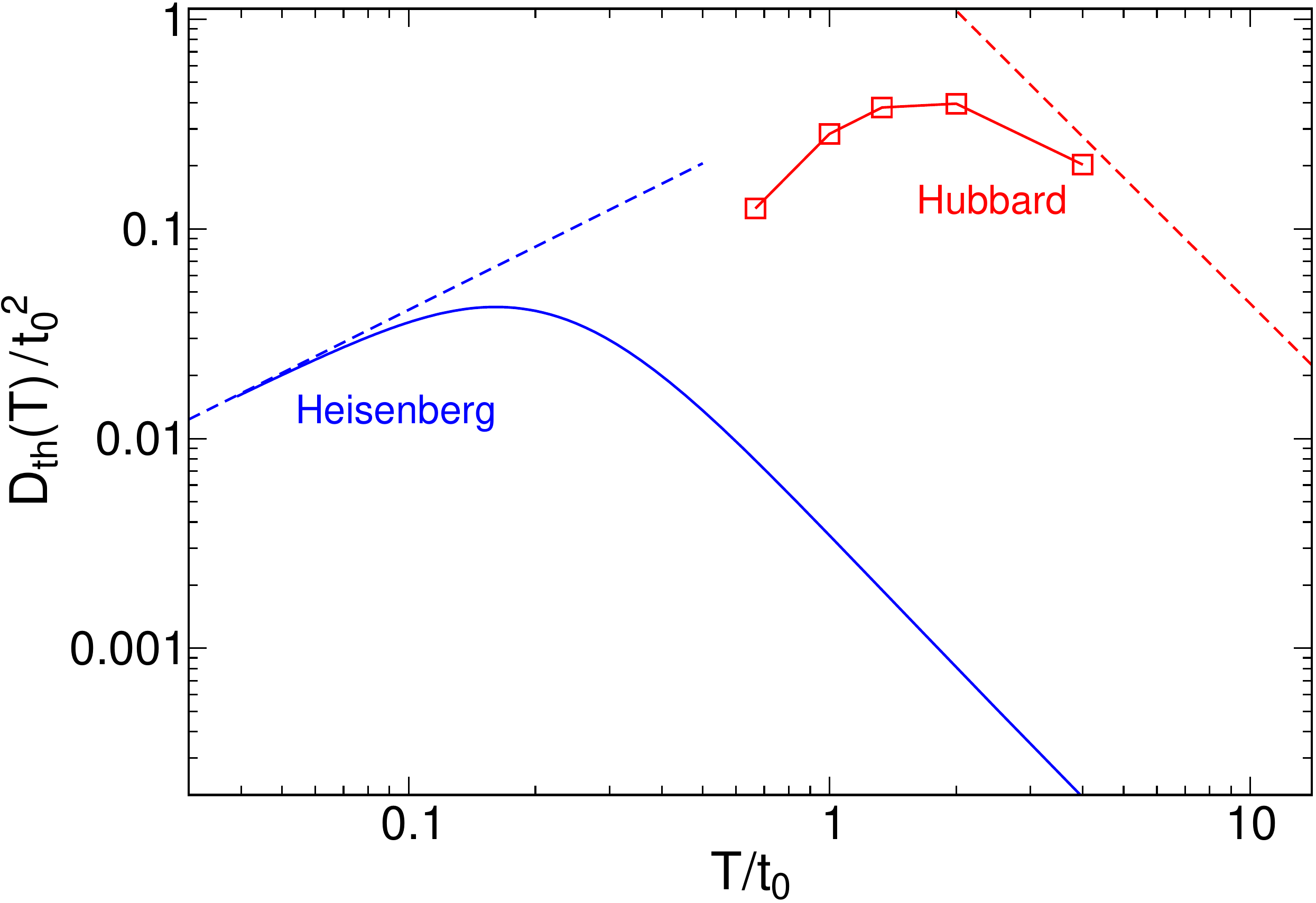}
\caption{(Color online) Drude weight of the Hubbard model at large $U/t_0=8$ in comparison with the result for an isotropic Heisenberg spin chain with an antiferromagnetic coupling $J=4t_0^2/U$. 
}\label{FigSupHeis}
\end{figure}

\begin{figure}[t]
\includegraphics[width=0.9\columnwidth]{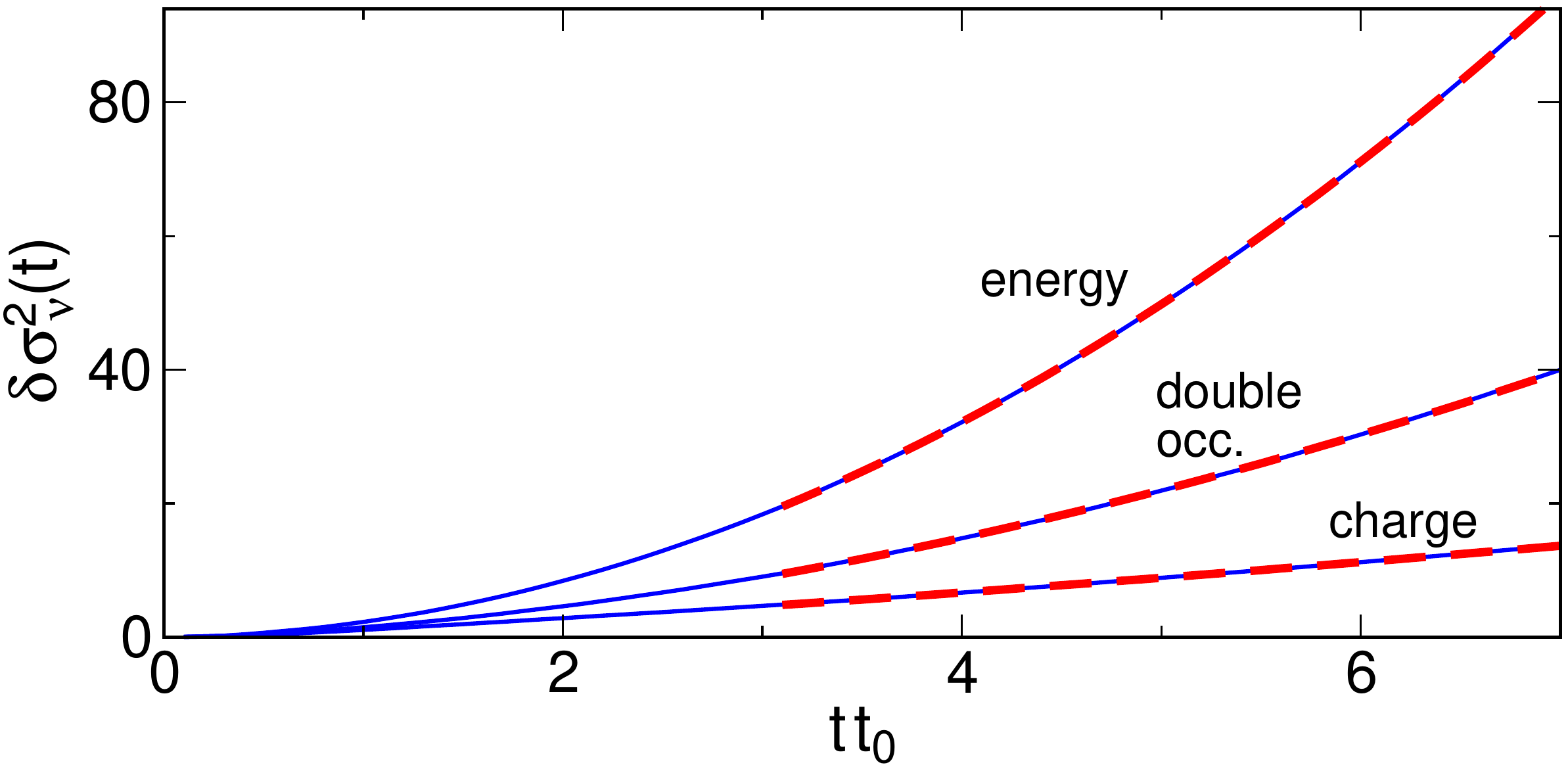}
\caption{(Color online) Variances $\delta\sigma^2_{\rm th,ch,d}(t)$ for the data shown in  Figs.~4(a-c) of the main text. The dashed lines show fits to $\delta\sigma^2_{\nu}(t)\sim t^\alpha$ ($\nu=\mbox{th,ch,d}$) for times $tt_0>3$. The fitted exponents read $\alpha_\tn{th}=1.96$, $\alpha_\tn{ch}=1.28$, and $\alpha_\tn{d}=1.76$.
}\label{FigSupVar}
\end{figure}

The thermal conductivity $\kappa$ is measured under the condition of $\langle I \rangle =0 $ and, therefore, is given by 
\begin{equation}
\kappa =  L_{22} - L_{21} L^{-1}_{11} L_{12}\,.\label{eq:lij}
\end{equation}
Alternatively, one can work with the currents $I_{\rm th}$ and $I$ and corresponding transport coefficients $M_{ij}$. The thermal conductivity 
is then given by
\begin{equation}
\kappa =  M_{22} - M_{21} M^{-1}_{11} M_{12}\,. \label{eq:mij}
\end{equation}
Thus, in principle, two terms contribute to $\kappa$, the second arises because of the thermoelectric coupling.
As is discussed in textbooks, Eqs.~\eqref{eq:lij} and \eqref{eq:mij} yield equivalent results for $\kappa$.

Let us first consider the corresponding Drude weights $D_{ij}$ associated with the $M_{ij}$ (in the notation of the main text, $D_{11}=D_{\rm s}$ and $D_{22}= D_{\rm th}$).
In the case of a half-filled system, e.g., the Mott insulator in 1D that is  considered in the main text, the energy and particle current 
have a different parity under particle-hole symmetry and therefore, 
the current-correlation function $C_{I,I_{\rm th}}(t) = \langle I(t) I_{\rm th} \rangle$  vanishes identically. As a consequence, $M_{12}= M_{21} =0 $ 
at half filling at all temperatures. This result is known from studies of, e.g., the thermopower which consequently also vanishes in 
a Mott insulator at $n=1$ in the presence of particle-hole symmetry \cite{beni75,peterson07}.
In other words, since we wrote down the Hamiltonian (3) in an explicitly particle-hole symmetric form for $V=0$, the chemical potential vanishes
at half filling. 
For the purpose of our work we conclude that
the Drude weight $D_{q}$ associated with the heat current $I_q$ 
is given by 
\begin{equation}
D_{q} = D_{\rm th}\,.
\end{equation}
Therefore, for the case studied in the main text, namely half filling, the thermal conductivity 
solely stems from the energy-current correlation function $C_{\rm th}(t) = \langle I_{\rm th}(t) I_{\rm th} \rangle$.

Note that the justification
of the Kubo formula for the thermal conductivity \cite{luttinger64} is more subtle than 
for  charge transport (see \cite{gemmer06} for a comprehensive discussion and additional references).
The main reason
is that a gradient in temperature is not a mechanical force  in the same sense as a voltage or gradient in chemical potential is. Therefore, 
one cannot add an additional term to the Hamiltonian that would contain the temperature gradient, which one would commonly treat as the perturbation
driving the system out of equilibrium.
An approach to circumvent that problem is to introduce so-called polarization operators (see, e.g., \cite{louis03} for a discussion).
Such polarization operators (essentially of the form $P=-\sum_l l h_l$) can then take the role of the external force. This construction,
however, requires open boundary conditions.
Other attempts make an assumption about local equilibrium 
or rely on the so-called  entropy production argument (see again \cite{gemmer06} for details).

Ref.~\cite{gemmer06} provides a derivation of the Kubo formula that does not rely on the assumption of local equilibrium.
That paper also attempts a direct comparison between time-dependent numerical solutions of certain models
in the presence of a temperature gradient and mostly confirms the validity of the Kubo formula. Some discrepancies were
observed in Ref.~\cite{gemmer06}, which, however, were later understood to be finite-size effects \cite{steinigeweg-comm}.

There is also substantial interest in extracting thermal currents and temperature profiles from simulations of open quantum
systems (see, e.g., \cite{michel05,prosen09,karevski09,arrachea09,prosen10,znidaric11,mendoza-arenas13,mendoza-arenas13a}),
where typically, master equations are solved. In some of these studies (see, e.g., \cite{znidaric11}), however, transport is {\it boundary}-driven, which realizes a different physical situation from
what underlies the derivation of Kuba formulae for charge, spin and energy transport, where the bulk experiences a gradient of some force.
As a consequence, such open quantum system simulations 
 can  yield qualitatively different results compared to the  Kubo formula, even for charge or spin transport (compare, e.g., \cite{znidaric11,karrasch14}).
Moreover, it is  important to keep in mind that conductivities are bulk properties by definition. Thus, any contact regions must be excluded in  measuring
currents and, in particular, temperature differences. Therefore, experiments commonly use a four-terminal setup to measure the thermal conductivity (see, e.g., \cite{sologubenko00, hess01,hess-diss}).
The technical challenge for such open quantum system simulations is therefore to be able to reach sufficiently large systems, as a result of
which a direct comparison to the Kubo formula is often not possible.

\subsection*{S3.~Comparison of DMRG results for different system sizes}
In Fig.~\ref{FigSupL}, we present DMRG results for $C_{\rm th}(t)$ for several different system sizes and temperatures at $U/t_0=8$ and $V=0$.
Since we are using open boundary conditions, finite-size effects manifest themselves by a drop of $C_{\rm th}(t)$ to zero and a sign change 
at a system-size dependent time $t^*(L)$. Very importantly, for $t<t^*(L)$, the data from different system sizes all coincide, implying that the results
are for the thermodynamic limit. Moreover, using data for open boundary conditions, one needs at least $L\gtrsim 20$ to get to times of the order of $4\sim 1/t_0$.
Using $L=100$, we clearly never reach $t^*$, hence the time-dependent data from such a large system is free of finite-size effects in the time window reached in the simulations. 
Note that the numerical effort is essentially independent of $L$ but primarily depends on the entanglement growth that limits the accessible times.

\subsection*{S4.~Comparison of the Drude weights of the FHM and Heisenberg chain}

In principle, two types of excitations should contribute to the thermal conductivity. First, for temperatures $T> \Delta_{\rm Mott} $, where $\Delta_{\rm Mott}$
is the Mott gap, charge excitations (i.e., excitations in the upper Hubbard band) become relevant. Second, at very low $T\ll \Delta_{\rm Mott}$, charge excitations
are frozen out and then gapless spin excitations should be the only available ones. Whether these two regimes can be resolved depends
on the ratio of the characteristic energy scale $J$ for magnetic excitations and the Mott gap $\Delta_{\rm Mott}$. 
In the large $U/t_0$ limit, $J=4 t_0^2/U$,  much smaller
than $\Delta_{\rm Mott} \sim U $. Therefore, the very low temperature dependence of $D_{\rm th}$ of the Hubbard model should be identical
to the one of the spin-1/2 Heisenberg model \cite{kluemper02,hm02}, where $D_{\rm th}$ increases linearly at low $T$, takes a maximum at $T\lesssim J/2$ and then decreases to zero with $1/T^2$.
As a consequence, the full $D_{\rm th}$ for large $U/t_0$ could have a double maximum structure and moreover, at $T\gg J$, the conductivity should solely stem from charge excitations since the spin system will effectively be at $T/J=\infty$ (This is often referred to as the spin-incoherent regime, see, e.g., \cite{feiguin11,Soltanieh-ha14}).
The Drude weights of the FHM at $U/t_0=8$ and the spin-1/2 Heisenberg chain \cite{kluemper02} are shown in Fig.~\ref{FigSupHeis}, confirming the
qualitative picture described above. Note also the difference in the maximum values of the Drude weights: the charge dominated $D_{\rm th}$
exceeds the Heisenberg contribution significantly.
The problem with numerically resolving the interesting crossover regime is that it requires a large $U$, which in the units of the FHM renders $J$  a
 very small temperature that at present cannot be reached.

\begin{figure}[t]
\includegraphics[width=0.7\columnwidth]{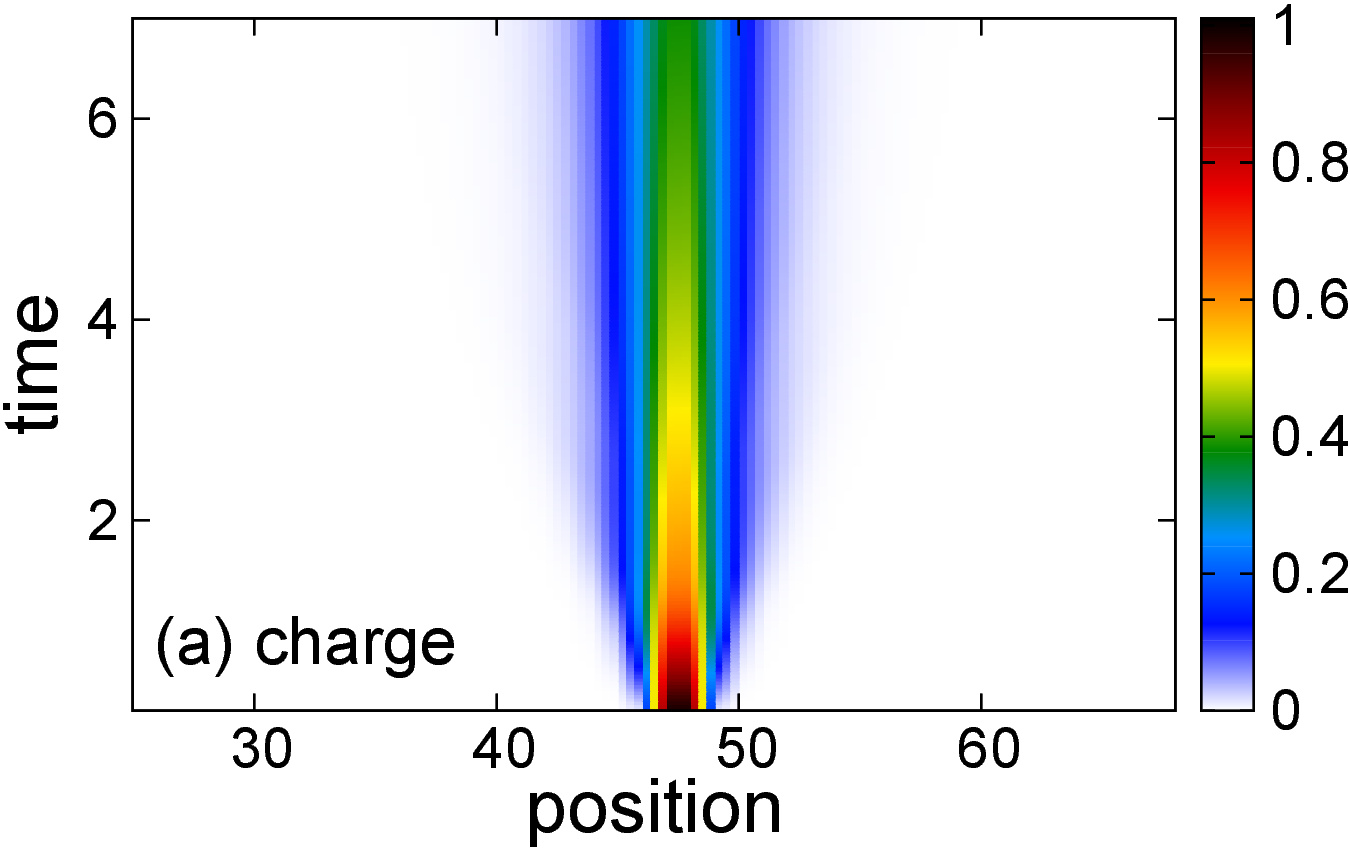}\vspace*{1ex}
\includegraphics[width=0.7\columnwidth]{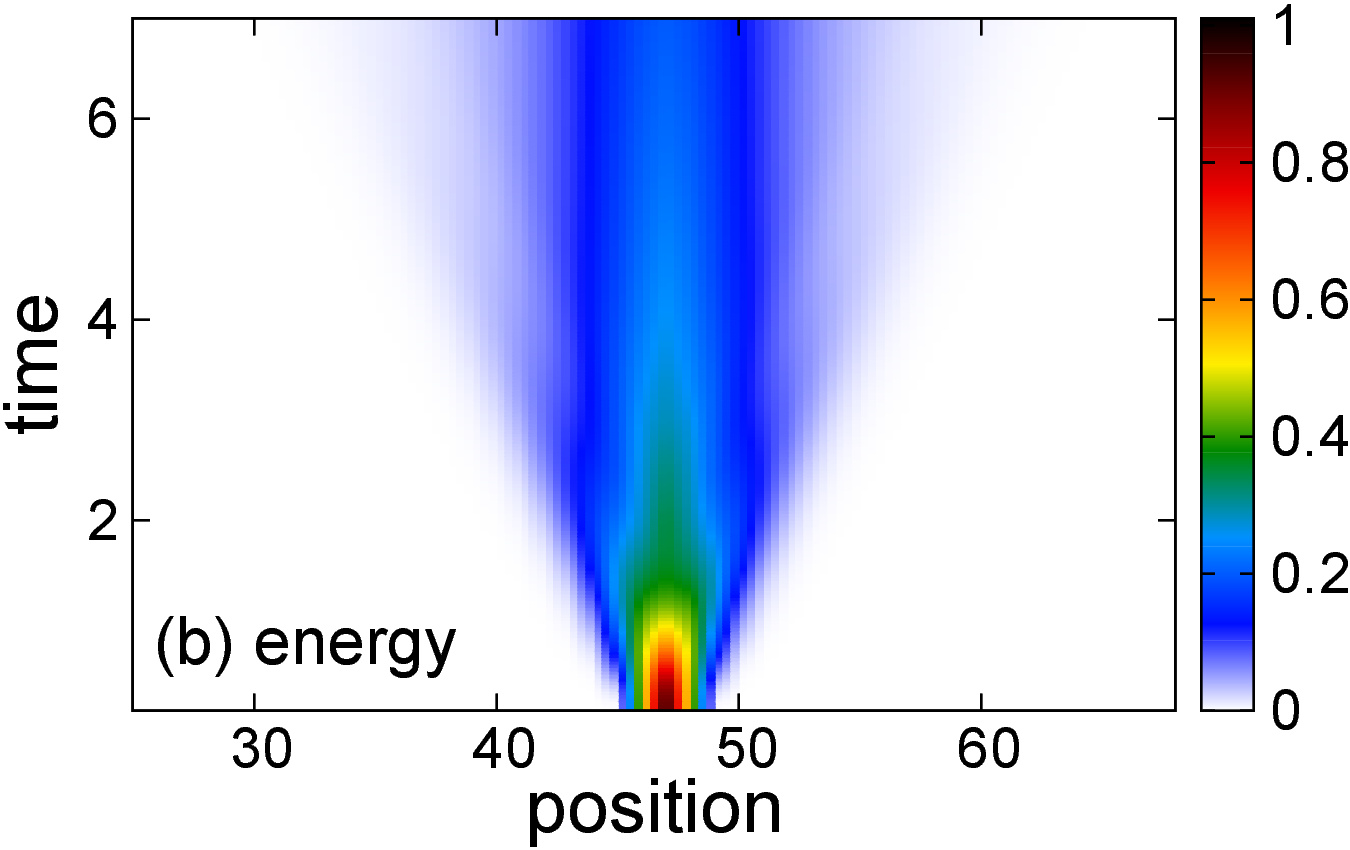}\vspace*{1ex}
\includegraphics[width=0.7\columnwidth]{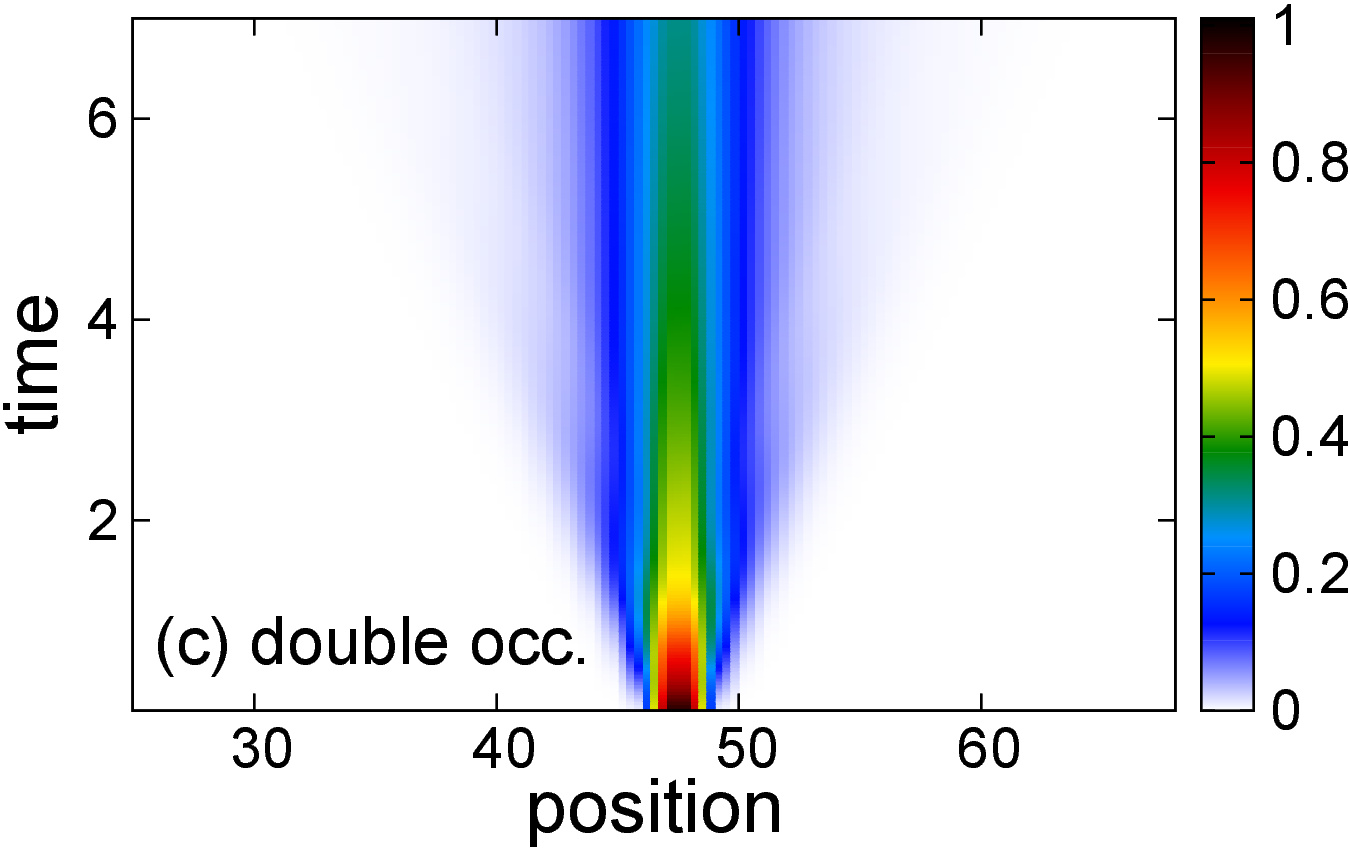}
\caption{(Color online) The same as in Figs.~4(a-c) of the main text, but for $V=U/4$.
The exponents of the time-dependence of the corresponding width $\delta \sigma_{\nu}^2 \propto t^{\alpha}$  ($\nu=\mbox{th,ch,d}$)
are $\alpha_\tn{th}=1.14$, $\alpha_\tn{ch}=0.72$, and $\alpha_\tn{d}=0.99$.
}\label{FigSupWavep}
\end{figure}

\phantom{\cite{ashcroft,steinigeweg-comm,karevski09,prosen10,znidaric11,mendoza-arenas13a,sologubenko00,hess01,feiguin11,Soltanieh-ha14}}

\bibliography{references}

\end{document}